\documentstyle[12pt,amssymb]{article}

\def\fnote#1#2{\begingroup\def\thefootnote{#1}\footnote{#2}\addtocounter
{footnote}{-1}\endgroup}

\def\inbar{\vrule height1.5ex width.4pt depth0pt}
\def\IB{\relax{\rm I\kern-.18em B}}
\def\IC{\relax\,\hbox{$\inbar\kern-.3em{\rm C}$}}
\def\ID{\relax{\rm I\kern-.18em D}}
\def\IE{\relax{\rm I\kern-.18em E}}
\def\IF{\relax{\rm I\kern-.18em F}}
\def\IG{\relax\,\hbox{$\inbar\kern-.3em{\rm G}$}}
\def\IH{\relax{\rm I\kern-.18em H}}
\def\II{\relax{\rm I\kern-.18em I}}
\def\IK{\relax{\rm I\kern-.18em K}}
\def\IL{\relax{\rm I\kern-.18em L}}
\def\IM{\relax{\rm I\kern-.18em M}}
\def\IN{\relax{\rm I\kern-.18em N}}
\def\IO{\relax\,\hbox{$\inbar\kern-.3em{\rm O}$}}
\def\IP{\relax{\rm I\kern-.18em P}}
\def\IQ{\relax\,\hbox{$\inbar\kern-.3em{\rm Q}$}}
\def\IR{\relax{\rm I\kern-.18em R}}
\def\IT{\relax{\rm I\kern-.18em T}}
\def\ZZ{\relax{\sf Z\kern-.4em Z}}

\def\a{\alpha}   \def\b{\beta}      
\def\e{\epsilon}      
     \def\si{\sigma}

\def\cA{{\cal A}} 
   \def\cF{{\cal F}}
 \def\cH{{\cal H}}  
   \def\cN{{\cal N}}
\def\cO{{\cal O}} \def\cP{{\cal P}}

\def\afrak{{\mathfrak a}} \def\ffrak{{\mathfrak f}}
\def\pfrak{{\mathfrak p}}
 \def\Hfrak{{\mathfrak H}}
\def\Ifrak{{\mathfrak I}} \def\Pfrak{{\mathfrak P}}

\def\bE{{\bar E}}

\def\bsi{\bar \sigma}

\def\IQmum{{\IQ(\mu_m)}} \def\IQmup{{\IQ(\mu_p)}}
 \def\ZZmum{{\ZZ[\mu_m]}}

\def\fnote#1#2{\begingroup\def\thefootnote{#1}\footnote{#2}\addtocounter
{footnote}{-1}\endgroup}
\def\beq{\begin{equation}}
\def\eeq{\end{equation}}
\def\bea{\begin{eqnarray}}
\def\eea{\end{eqnarray}}

\def\lleq#1{\label{#1}\eeq}

\def\notin{\ \hbox{{$\in$}\kern-.51em\hbox{/}}}

\def\lra{\longrightarrow}

  \def\E1Fq{E_1/\IF_q} 
  
\def\XFp{{X/\IF_p}}  \def\XFpr{{X/\IF_{p^r}}}

\def\IFqr{{\IF_{q^r}}}

\def\rmH{{\rm H}} 
\def\rmN{{\rm N}}

\def\rmdet{{\rm det}} \def\rmdim{{\rm dim}}
 \def\rmHom{{\rm Hom}}
\def\rmgal{{\rm Gal}}  \def\rmmod{{\rm mod}}
 \def\rmtr{{\rm tr}}

 \def\rmCl{{\rm Cl}}

\begin{document}
\parindent=0pt
\hfill {\bf NSF$-$ITP$-$01$-$158}

 \vskip 1.2truein

\centerline{\large {\bf  ARITHMETIC OF CALABI-YAU VARIETIES}}
\vskip .1truein

\centerline{\bf AND}

\vskip .1truein

\centerline{\large {\bf RATIONAL CONFORMAL FIELD THEORY}}

\vskip 0.6truein

\centerline{\sc Rolf Schimmrigk\fnote{\diamond}{email:
netahu@yahoo.com, rks@canes.gsw.edu}}

\vskip .2truein

\centerline{\it Georgia Southwestern State University}\vskip
.05truein \centerline{\it 800 Wheatley St, Americus, GA 31709}

\vskip 1.1truein

\baselineskip=19pt

\centerline{\bf ABSTRACT:} \vskip .2truein
 \noindent It is proposed that certain techniques from
 arithmetic algebraic geometry provide a framework which is useful
 to formulate a direct and intrinsic link between the
 geometry of Calabi-Yau manifolds and the underlying conformal field
 theory. Specifically it is pointed out how the algebraic number field
 determined by the fusion rules of the conformal field theory
 can be derived from the number theoretic structure of
 the cohomological Hasse-Weil L-function determined by
  Artin's congruent zeta function of the algebraic variety.
In this context a natural number theoretic characterization arises
for the quantum dimensions in this geometrically determined
algebraic number field.

\renewcommand\thepage{}
\newpage

\baselineskip=22pt
\parskip=.2truein
\parindent=0pt
\pagenumbering{arabic}

\section{Introduction}

In a stunning development Gepner suggested in 1987 that Calabi-Yau
varieties are closely related to particular types of
two-dimensional conformal field theories with integral central
charges \cite{g87}. For the simplest class of models defined by
weighted projective Brieskorn-Pham threefold varieties this
relation might be formulated in a preliminary way as follows \beq
\IP_{(k_1,k_2,k_4,k_4,k_5)}\left[d\right]~ \ni ~ X=\left\{p(z_i) =
\sum_i z_i^{d/k_i}=0\right\} ~ \sim ~ \bigotimes_{i=1}^5 ~ {\rm
SU(2)}_{\frac{d}{k_i}-2}. \lleq{BPvar} Here the levels
$\ell=\left(\frac{d}{k_i} -2\right)$ of the SU(2)$_{\ell}$
conformal theories on the rhs are determined by the weights $k_i$
of the weighted projective space, denoted by
$\IP_{(k_1,...,k_5)}$, and the degree $d:=\sum_i k_i$ of the
defining polynomial of the variety $X$. For Brieskorn-Pham
varieties the quotients $d/k_i$ are natural numbers. On the
conformal field theory side we implicitly assume the GSO
projection and that the affine characters of the partition
function of the individual conformal factors are chosen to be the
diagonal ones. The complete class of these models has been
determined in \cite{ls90, fkss90}.

Gepner provided support for this relation by computing the
massless spectrum of his theories and by showing that in two
instances they agreed with the cohomology groups of known
Calabi-Yau threefolds \cite{chsw85, s87}. Crucial insight into
this surprising relation was obtained by Martinec \cite{m89} and
Lerche, Vafa, and Warner \cite{vw89,lvw89} who noted that an
intermediate framework is provided by the concept of a
Landau-Ginzburg theory. The latter defines a mean-field theory of
the exactly solvable tensor model. In particular the important
paper \cite{v88} by Vafa illuminated both the computation of the
conformal field theory spectrum as well as providing effective
insight into the cohomology of algebraic variety. Vafa's insight
in particular made it possible to systematically investigate the
relation between conformal field theory spectra and the cohomology
of algebraic varieties and to extend this comparison to the
complete class of models including non-diagonal affine invariants
of the partition functions \cite{ls90,fkss90}. Later more detailed
insight into the nature of this relation via the intermediary
Landau-Ginzburg theory was obtained by Witten \cite{w92}.

At present, however, neither a direct nor a mathematically
rigorous framework exists which allows to establish relations
between algebraic varieties and conformal field theories.
Furthermore the Landau-Ginzburg framework, important as it has
been, is incomplete and does not allow to derive key ingredients
of the conformal field theory. One might instead hope for a
framework in which it is possible to derive the essential
ingredients of the conformal field theory directly from the
algebraic variety itself, without the intermediate Landau-Ginzburg
formulation. A priori it might appear unlikely that such an
approach exists because numbers associated to algebraic varieties
are usually integers (such as dimension, cohomology dimensions and
indices associated to complexes), whereas the numbers appearing in
the underlying exactly solvable conformal field theory are usually
rational numbers (such as the central charge and the anomalous
dimensions).

It is the purpose of this article to suggest that an intrinsic,
direct, and mathematically rigorous framework which allows to
derive certain conformal field theoretic quantities directly from
the algebraic Calabi-Yau variety is provided by arithmetic
algebraic geometry. In this framework algebraic varieties $X$ are
defined not over a continuous field, like the real numbers $\IR$,
or the complex number field $\IC$, but over discrete finite
fields, denoted by $\IF_q$, where $q \in \IN$ denotes the number
of elements of the field. The particular field of reduction of the
variety is specified by writing $X/\IF_q$, leading to a reduced
variety consisting of a finite number of points.

The choice of any finite $q$ would appear to be arbitrary, and
physically ill-motivated. For small $q$ the field $\IF_q$ would
define a large scale lattice structure which one might expect to
provide only rough information about the structure of the variety.
More sensible would be to consider an infinite sequence of ever
larger finite fields which probe the variety at ever smaller
scales. It is this consideration which leads us to the concept of
counting the number of solutions $N_{q^r}=\#(X/\IF_{q^r})$ and to
ask what, if any, interesting information is provided by the
numbers determined by the extensions $\IFqr$ of degree $r$ of the
finite field $\IF_q$. The following paragraphs outline the
strategy envisioned in this program and briefly introduce the key
ingredients used in this paper.

The starting point of the arithmetic considerations in Section 3
is to arrange the sequence of reductions of the variety over the
finite fields $\IF_{q^r}$ into a useful form. This can be achieved
via Artin's congruent zeta function, essentially defining an
exponential sum of a generating function constructed from the
numbers $N_{q^r}$. For reasons just described we should not
restrict the construction to local considerations at particular
integers. Hence we need some way to pass to a global description
in which all reference to fixed scales has been erased. This
leads, in Section 4, to the concept of the global Hasse-Weil
L-function of a Calabi-Yau variety. This L-function will collect
the information at all rational primes. This step can be taken
because the Weil conjectures show that in the present context
Artin's zeta function is a rational function determined by the
cohomology of the variety. The Hasse-Weil L-function therefore is
a cohomological L-function.

At this point the only number theoretic algebraic structures that
have appeared are the finite fields $\IF_q$ and their extensions.
A brief review of some of the aspects of the underlying conformal
field theoretic aspects, provided in Section 2, reveals that this
is not enough. As mentioned above, the structures that enter in
Gepner's construction and its generalizations are rational, i.e.
they lead to central charges and spectra of anomalous dimensions
which are rational numbers. It is not clear how to recover these
numbers from the intrinsic geometry of the variety. It turns out
that it is more useful in the present context to encode the
conformal field theoretic  information in an alternative way by
mapping the central charge and the anomalous dimensions via the
Rogers dilogarithm into the quantum dimensions associated to the
physical fields. These (generalized) quantum dimensions are
elements of certain algebraic number fields which are determined
by the fusion rules of the conformal field theory.

This suggests that we consider the structure of algebraic number
fields in more detail. It is in this context that Hecke introduced
a general notion of L-functions which are determined completely by
the prime ideals of the number field, generalizing earlier results
obtained by Dirichlet and Dedekind. Such number field L-functions
essentially count the inverse of the norms of ideals in the number
field weighted by a character of this field.

The surprise now is that the cohomological Hasse-Weil type
L-function of the Calabi-Yau variety in fact happens to be
determined by a number field L-function of the type introduced by
Hecke. This is the link that allows us to recover directly from
the intrinsic geometry of the variety the number theoretic
framework relevant to the underlying conformal field theory. Once
the fusion field has been identified one can then further explore
the arithmetic r\^{o}le played by the quantum dimensions in this
field.

The strategy to be employed in this paper thus can be summarized
as follows. First consider the arithmetic structure of Calabi-Yau
varieties via Artin's congruent zeta function. Next define the
global cohomological Hasse-Weil L-function via local factors from
the congruent zeta function. Finally interpret the Hasse-Weil
L-series as Hecke L-series of an algebraic number field. In this
way we can recover the algebraic fusion field of the conformal
field theory from the algebraic variety.

The program to use arithmetic geometry to illuminate the conformal
field theoretic structure of Calabi-Yau varieties originated
several years ago. The idea initially was to develop further some
results derived by Bloch and Schoen in support of the
Beilinson-Bloch conjectures and apply them in the context of the
conformal field theory/Calabi-Yau relation \cite{s95}. More
recently arithmetic considerations have been discussed in
different contexts in refs. \cite{m98} and \cite{cdv00}.

\section{Exactly solvable Calabi-Yau manifolds}

In the following we investigate the simplest class among weighted
Calabi--Yau hypersurfaces described by higher-dimensional analogs
of polynomials of Brieskorn--Pham type. These manifolds are
distinguished by the fact that for every members there exists a
particular point in moduli space for which the underlying exactly
solvable conformal field theory is known explicitly. As indicated
in (\ref{BPvar}) the varieties are described as zero-loci of
polynomials in weighted projective space.

 The underlying rational conformal field theories
suggested by Gepner are exactly solvable tensor products of N$=$2
superconformal minimal models of central charge \beq
c=\frac{3k}{k+2}.\eeq  These factors are glued together by
Gepner's construction \cite{g87}, to which we refer for a more
detailed discussion of the conformal field theory aspects.  The
physical string theoretic spectrum of the theory is determined by
the conformal field theory chiral primary operators fields with
integral anomalous dimension. These fields are constructed from
operators in the individual SU(2)$_{k}$ factors with the anomalous
dimensions \beq \Delta_j^{(k)} =
\frac{j(j+2)}{4(k+2)},~~~,j=0,...,k. \eeq The SU(2) factors are
the nontrivial interacting conformal field theories that provide
the building blocks of the Gepner models.

It would thus appear that a possible, and perhaps natural,
strategy should be an attempt to recover these rational numbers
given by the underlying conformal field theory from the intrinsic
geometry of the algebraic variety.  It turns out however that it
is more useful to proceed along a more indirect route in which the
information of the anomalous dimensions and the central charge is
first encoded in a set of numbers which are not elements of $\IQ$,
i.e. they are not rational. This will provide deeper information
about the conformal field theory and is the route taken in this
paper.

In order to establish this relation recall the modular $S$--matrix
of the affine theories. Consider maps
$$ \chi: \Hfrak \times \IC \lra \IC$$
defined on the upper half plane $\Hfrak$ and $\IC$ parametrizing
the charges of the extended theory
$$ \chi_i(\tau, u):= \rmtr_{\cH_i} q^{L_0 - \frac{c}{24}} e^{2\pi i uJ_0}.
$$
In the simple case of SU(2)$_k$ the modular transformations
$$ \chi_i\left(-\frac{1}{\tau}, \frac{u}{\tau}\right) =
e^{\pi i ku^2/2} \sum_j S_{ij} \chi_j(\tau, u)$$ lead to the
modular $S$-matrix
 \beq S_{ij} =
\sqrt{\frac{2}{k+2}}~~\sin\left(\frac{(i+1)(j+1)\pi}{k+2}\right),~~~~~0\leq
i,j \leq k. \eeq

With these matrices one can define the generalized quantum
dimensions as \beq Q_{ij} = \frac{S_{ij}}{S_{0j}} \eeq for the
SU(2) theory at level $k$. The importance of these numbers derives
from the fact that even though they do not directly provide the
scaling behavior of the correlation functions, they do contain the
complete information about the anomalous dimensions as well as the
central charge
 \beq \frac{1}{L(1)} \sum_{i=1}^k
L\left(\frac{1}{Q_{ij}^2}\right) = \frac{3k}{k+2} - 24
\Delta_j^{(k)} +6j, \lleq{nrt} where $L$ is Rogers' dilogarithm
\beq L(z) = Li_2(z) +{\small \frac{1}{2}} \log(z)~\log(1-z) \eeq
and $Li_2$ is Euler's classical dilogarithm \beq Li_2(z) =
\sum_{n\in \IN} \frac{z^n}{n^2}.\eeq It follows that the quantum
dimensions contain the essential information about the spectrum of
the conformal field theory and Rogers' dilogarithm provides, via
Euler's dilogarithm, the map from the quantum dimensions to the
central charge and the anomalous dimensions. A review of these
results and references to the original literature can be found in
\cite{ak94}.

It is possible to gain insight into the nature of these particular
numbers.  The point here is that the extended quantum dimensions
$Q_{ij}$ are elements of a particular field which extends the
field of rational numbers. To determine the structure of this
field one can proceed as follows. The starting point is the fusion
algebra which describes a product of the chiral primary fields of
the conformal field theory \beq \Phi_i \star \Phi_j = (\cN_i)_j^k
\Phi_k, \eeq where the resulting coefficients $(\cN_i)_j^k$ define
the so-called fusion matrices $\cN_i$. The generalized quantum
dimensions are eigenvalues of these fusion matrices and therefore
they are solutions of the characteristic polynomials \beq
det\left(Q \cdot {\bf 1} - \cN_i\right) =0. \eeq Because the
$\cN_i$ are integral matrices the characteristic polynomial has
integral coefficients and it is also normalized, i.e. its leading
coefficient is unity. Therefore the $Q_{ij}$ are algebraic
integers in some algebraic number field $K$ over the rational
numbers $\IQ$. An algebraic number field $K$ of degree $n=[K:\IQ]$
is defined as the set of solutions of an equation \beq a_0 + a_1x
+ a_2x^2 + \cdots + a_nx^n=0, \lleq{algnum} with coefficients
$a_i$ which are rational integers $a_i \in \ZZ$. The set of
algebraic integers of $K$ is defined as the set of solutions of
equation (\ref{algnum}) whose leading coefficient is unity
$a_n=1$. The resulting expressions in general involve rational
coefficients.

The extension $K/\IQ$ is normal and because $\IQ$ is of
characteristic zero this extension is a Galois extension, i.e. its
Galois group Gal(K/\IQ) is abelian. Hence it follows from the
theorem of Kronecker and Weber that the field $K$ is contained in
some cyclotomic field $\IQmum$ where $\mu_m$ is the cyclic group
of order $m$, generated by the primitive $m$'th root of unity
\cite{dg91}. We will call the minimal cyclotomic field which
contains all the quantum dimensions the fusion field of the
rational conformal field theory.

 In the present case we can be more specific: the generalized
quantum dimensions take values  $Q_{ij}\left( {\rm SU(2)}_k
\right) \in \IQ(\mu_{2(k+2)})$ and because the quantum dimensions
are real they are elements of a real subfield of the cyclotomic
field. For the conformal field theory underlying the quintic
hypersurface one finds e.g.
 \beq Q_i\left({\rm SU(2)}_3\right)
  \in \left\{1,\frac{\tiny 1}{\tiny 2}(1+\sqrt{5})\right\}
  \subset ~\IQ(\sqrt{5})
\eeq i.e. the fusion field of the affine conformal field theory
underlying the quintic is the cyclotomic field $\IQ(\mu_5)$.

The problem of recovering the conformal field theory spectrum
encoded by the anomalous dimensions can now be rephrased into the
question whether it is possible to derive the fusion field of the
quantum dimensions from the arithmetic structure of the
corresponding Calabi--Yau manifold. Once this is achieved one can
ask for an intrinsic field theoretic interpretation of the real
quantum dimensions within the fusion field.

\section{Artin's Congruent Zeta Function of Calabi-Yau threefolds}

The starting point of the arithmetic analysis is the set of
 Weil conjectures \cite{w49}, the proof of which was completed by
 Deligne \cite{d74}.
For algebraic varieties the Weil--Deligne result states a number
of structural properties for the congruent zeta function at a
prime number $p$ defined as \beq Z(\XFp, t) \equiv
exp\left(\sum_{r\in \IN} \# \left(\XFpr\right)
\frac{t^r}{r}\right). \eeq The motivation to arrange the numbers
$N_{p,r}= \# \left(\XFpr\right)$ in this unusual way originates
from the fact that they often show a simple behavior, as a result
of which the zeta function can be shown to be a rational function.
This was first shown by Artin in the 1920s for hyperelliptic
function fields. Further experience by Hasse, Weil, and others led
to the conjecture that this phenomenon is more general,
culminating in the Weil conjectures.

These general claims can be summarized as follows.
\begin{enumerate}
\item The zeta function of an algebraic variety of dimension $d$
satisfies a functional equation \cite{k94} \beq
Z\left(\XFp,\frac{1}{p^dt}\right) = (-1)^{\chi +\mu} p^{d\chi/2}
t^{\chi} Z(\XFp,t),\eeq where $\chi$ is the Euler number of the
variety $X$ over the complex numbers $\IC$. Furthermore $\mu$ is
zero when the dimension $d$ of the variety is odd, and $\mu$ is
the multiplicity of $-p^{d/2}$ as an eigenvalue of the action
induced on the cohomology by the Frobenius automorphism $\Phi:
X\lra X,~x\mapsto x^p$.
\item $ Z(\XFp,t)$ is a rational function which can be written as
\beq  Z(\XFp,t)=\frac{\prod_{j=1}^d
\cP^{(p)}_{2j-1}(t)}{\prod_{j=0}^d \cP^{(p)}_{2j}(t)}, \eeq where
\beq \cP_0^{(p)}(t)=1-t,~~~ \cP_{2d}^{(p)}(t)=1-p^dt \eeq and for
$1\leq i \leq 2d-1$ \beq \cP_i^{(p)}(t) = \prod_{j=1}^{b_i}
\left(1-\beta^{(i)}_j(p) t\right), \eeq with algebraic integers
$\b^{(i)}_j(p)$. The $b_i$ denote the Betti numbers of the
variety, $b_i={\rm dim~ H}^i_{\rm DeRham}(X)$. The rationality of
the zeta function was first shown by Dwork \cite{bd60} by adelic
methods.
\item The algebraic integers $\beta^{(i)}_j(p)$
satisfy the Riemann hypothesis \beq |\beta^{(i)}_j(p)| =
p^{i/2},~~~~\forall i.\eeq It is this part of the Weil conjectures
which resisted the longest, and was finally proved by Deligne.
\end{enumerate}

We are interested in Calabi--Yau threefolds with finite
fundamental group, i.e. Calabi-Yau varieties with
$h^{1,0}=0=h^{2,0}$.  For such varieties the expressions above
simplify considerably.  For an arbitrary Calabi-Yau threefold the
congruent zeta function takes the form \beq Z(\XFp, t)=
\frac{\cP^{(p)}_3(t)}{(1-t)\cP^{(p)}_2(t)\cP^{(p)}_4(t)(1-p^3t)}
\eeq with \beq deg(\cP^{(p)}_3(t)) = 2+2h^{(2,1)},~~~
deg(\cP^{(p)}_2(t)) = h^{(1,1)}. \eeq This follows from the fact
that for non-toroidal Calabi-Yau threefolds we have $b_1=0$. In
general the indicated polynomials $\cP_i^{(q)}(t)$ left
unspecified are non-trivial for the Brieskorn-Pham varieties of
interest in the present context. This nontrivial structure arises
from the fact that weighted projective varieties admit
singularities which have to be resolved and therefore lead to
contributions of $H^2(X)$.

 The relation between the Calabi-Yau
varieties and conformal field theories is universal for all
members of the Brieskorn-Pham type, independent on whether they
had to be resolved or were smooth initially. This suggests that it
must be possible to recover the basic ingredients necessary to
derive the conformal field theory already from the class of smooth
manifolds. For varieties which do not have to be resolved the
structure of the zeta functions is particularly simple. In
particular one finds that the congruent zeta function of smooth
Calabi-Yau hypersurfaces in weighted projective fourspace is given
by
 \beq Z(\XFp, t)=
\frac{\cP^{(p)}_3(t)}{(1-t)(1-pt)(1-p^2t)(1-p^3t)} \eeq with \beq
deg(\cP^{(p)}_3(t)) = 2+2h^{(2,1)}\eeq and \beq \sum_{i=1}^{b_3}
\beta_i^{(3)} = 1+p+p^2+p^3 -\#(\XFp). \lleq{zetacysmooth} Hence
for smooth hypersurface threefolds the only interesting
information of the zeta function is encoded in the polynomials
$\cP_3^{(p)}(t)$.

\section{Hasse-Weil L-function}

 We see from the rationality of the zeta function that the basic information
 of this quantity is parametrized by the cohomology of the variety.
 More precisely, one can show that  the $i$'th polynomial $\cP_i^{(p)}(t)$ is
 associated to the action induced by the Frobenius morphism on the
 $i$'th cohomology group $\rmH^i(X)$. In order to gain insight
 into the arithmetic information encoded in this Frobenius action it
 is useful to decompose the zeta function. This leads to the
 concept of a local L-function that is associated to the
 polynomials $\cP_i^{(p)}(t)$ via the following definition.

 Let $\cP_i^{(p)}(t)$ be the polynomials
 determined by the rational congruent zeta function over the field
 $\IF_p$. The $i$'th L-function of the variety
 $X$ over $\IF_p$ then is defined via \beq L^{(i)}(X/\IF_p, s) =
 \frac{1}{\cP_i^{(p)}(p^{-s})}.\eeq

 Such L-functions are of interest for a number reasons. One of
these is that often they can be modified by simple factors so that
after analytic continuation they (are conjectured to) satisfy some
type of functional equation.

It was mentioned in the previous section that the geometry/CFT
relation must hold for the simplest type of varieties, in
particular those that do not need any kind of resolution. When
considering the cohomology of such simple varieties in dimensions
one through four it becomes clear that independent of the
dimension of the Calabi-Yau variety the essential ingredient is
provided by the cycles that span the middle-dimensional
(co)homology. Hence even though in general the cohomology can be
fairly complex, in particular for Calabi-Yau fourfolds,  the  only
local factor in the L-function that is relevant for the present
discussion is $\cP_d^{(p)}(t)$ for $d=\rmdim_{\IC}X$.

This suggests a natural generalization of the concept of the
Hasse-Weil L-function of an elliptic curve.
 Let $X$ be a Calabi-Yau $d$-fold with $h^{i,0}=0$ for $0<i<n-1$
 and denote by $P(X)$ the set of good prime numbers of $X$, i.e.
 those prime numbers for which the variety has good reduction over
 $\IF_p$.
Then its associated Hasse-Weil L-function is defined as \beq
L_{\rm HW}(X,s) = \prod_{p\in P(X)}
\frac{1}{\cP^{(p)}_d\left(p^{-s}\right)} = \prod_{p\in P(X)}
\frac{1}{\prod_{j=1}^{b_3} \left(1-\beta_j^{(d)}(p)p^{-s}\right)}.
\eeq  The restriction to good primes is a nontrivial requirement.
Even though the Brieskorn-Pham type varieties are all smooth (i.e.
transverse) as varieties over the complex numbers $\IC$ they do
not retain this property for arbitrary subfields $K\subset \IC$.
In the case of the quintic considered below the prime $p=5$ is an
example of a bad prime. The existence of bad primes complicates
the whole theory considerably, but for our purposes we can ignore
the additional factors of the completed L-functions that are
induced by these bad primes.

As mentioned above, for smooth weighted CY hypersurfaces the
Hasse-Weil L-function then contains the complete arithmetic
information of the congruent zeta function. For reasons that will
become clear below it is of importance that the above Hasse-Weil
function can be related to a Hecke L-function, induced by Hecke
characters. The main virtue of such characters is that as the
simpler Dirichlet characters they are multiplicative maps. It is
this multiplicativity which is of essence for the present
framework.

\subsection{The Quintic Example}

Consider the Calabi-Yau variety defined by the quintic
hypersurface in ordinary projective fourspace $\IP_4$. We denote
the general $h^{2,1}=101$ complex dimensional family of quintic
hypersurfaces in projective fourspace $\IP_4$ by $\IP_4[5]$ and
consider the threefold defined by \beq \IP_4[5] \ni X=
\left\{\sum_{i=0}^4 x_i^5=0\right\}.\eeq It follows from
Lefshetz's hyperplane theorem that the cohomology below the middle
dimension is inherited from the ambient space. Thus we have
$h^{1,0}=0=h^{0,1}$ and $h^{1,1}=1$ while $h^{2,1}=101$ follows
from counting monomials of degree five. Following Weil \cite{w49}
the zeta function is determined by (\ref{zetacysmooth}), where the
numerator is given by the polynomial
$\cP_3^{(p)}(t)=\prod_{i=1}^{204} (1-\beta_i^{(3)}(p)t)$ which
takes the form  \beq \cP_3^{(p)}(t)= \prod_{\a \in \cA}
\left(1-j_p(\a) t \right).\eeq This expression involves the
following ingredients. Define $\ell = (5,p-1)$ and rational
numbers $\a_i$ via $\ell \a_i\equiv 0(\rmmod~1)$. The set $\cA$ is
defined as \beq \cA=\{\a=(\a_0,...,\a_4) ~|~0<\a_i <1,~5\a_i\equiv
0(\rmmod~1), \sum_i \a_i=0(\rmmod~1)\}.\eeq Defining the
characters $\chi_{\a_i}\in \hat{\IF_p}$ in the dual of $\IF_p$ as
$\chi_{\a_i}(u_i)=exp(2\pi i\a_i s_i)$ with $u_i=g^{s_i}$ for a
generating element $g \in \IF_p$, the factor $j_p(\a)$ finally is
determined as \beq j_p(\a) =\frac{1}{p-1}\sum_{\sum_i u_i=0}
\prod_{i=0}^4 \chi_{\a_i}(u_i).\eeq

We thus see that the congruent zeta function leads to the
Hasse-Weil L-function associated to a Calabi-Yau threefold \beq
L_{\rm HW}(X,s) = \prod_{p \in P(X)} \prod_{\a \in \cA}
\left(1-\frac{j_p(\a)}{p^s}\right)^{-1},\eeq ignoring the bad
primes, which are irrelevant for our purposes.

\section{L-Functions and Algebraic Number Fields}

As mentioned in the Introduction, the surprising aspect of the
Hasse-Weil L-function is that it is determined by another, a
priori completely different kind of L-function that is derived not
from a variety but from a number field. It is this possibility to
interpret the cohomological Hasse-Weil L-function as a field
theoretic L-function which establishes the connection that allows
to derive number fields $K$ from algebraic varieties $X$.

For the case at hand the type of L-function that is relevant is
that of a Hecke L-function determined by a Hecke character, more
precisely an algebraic Hecke character. Following Weil we will see
that the relevant field for our case is the extension $\IQmum$ of
the rational integers $\IQ$ by roots of unity, generated by
$\xi=e^{2\pi i/m}$ for some rational integer $m$. It turns out
that these fields fit in very nicely with the conformal field
theory point of view. In order to see how this works this Section
first describes the concept of Hecke characters and then explains
how the L-function fits into this framework.

There are many different different definitions of algebraic Hecke
characters, depending on the precise number theoretic framework.
Originally this concept was introduced by Hecke \cite{h18} as
Gr\"ossencharaktere of an arbitrary algebraic number field. In the
following Deligne's adaptation of Weil's Gr\"ossencharaktere of
type $A_0$ is used \cite{d77}. Let $\cO_K \subset K$ be the ring
of integers of the number field $K$, $\ffrak \subset \cO_K$ an
integral ideal, and $E$ a field of characteristic zero. Denote by
$\Ifrak_{\ffrak}(K)$ the set of fractional ideals of $K$ that are
prime to $\ffrak$ and denote by $\Pfrak_{\ffrak}(K)$ the principal
ideals $(\a)$ of $K$ for which $\a\equiv 1(\rmmod~\ffrak)$. An
algebraic Hecke character modulo $\ffrak$ is a multiplicative
function $\chi$ defined on the ideals $ \Ifrak_{\ffrak}(K)$ for
which the following condition holds. There exists an element in
the integral group ring $\sum n_{\si}\si \in \ZZ[\rmHom(K,\bE)]$,
where $\bE$ is the algebraic closure of $E$, such that if $(\a)
\in \Pfrak_{\ffrak}(K)$ then \beq \chi((\a)) = \prod_{\si}
\si(\a)^{n_{\si}}. \eeq Furthermore there is an integer $w>0$ such
that $n_{\si}+n_{\bsi}=w$ for all $\si \in \rmHom(K,\bE)$. This
integer $w$ is called the weight of the character $\chi$.

Given any such character $\chi$ defined on the ideals of the
algebraic number field $K$ we can follow Hecke and consider a
generalization of the Dirichlet series via the L-function \beq
L(\chi,s)=\prod_{\stackrel{\pfrak \subset \cO_K}{\pfrak ~{\rm
prime}}} \frac{1}{1-\frac{\chi(\pfrak)}{\rmN\pfrak^s}} =
\sum_{\afrak \subset \cO_K} \frac{\chi(\afrak)}{\rmN\afrak^s},\eeq
where the sum runs through all the ideals. Here $\rmN\pfrak$
denotes the norm of the ideal $\pfrak$, which is defined as the
number of elements in $\cO_K/\pfrak$. The norm is a multiplicative
function, hence can be extended to all ideals via the prime ideal
decomposition of a general ideal. If we can deduce from the
Hasse-Weil L-function the particular Hecke character(s) involved
we will be able to derive directly from the variety in an
intrinsic way distinguished number field(s) $K$.

Insight into the nature of number fields can be gained by
recognizing that for certain extensions $K$ of the rational number
$\IQ$ the higher Legendre symbols provide the characters that
enter the discussion above. Inspection then suggests that we
consider the power residue symbols of cyclotomic fields $K=\IQmum$
with integer ring $\cO_K=\ZZmum$. The transition from the
cyclotomic field to the finite fields is provided by the character
which is determined for any algebraic integer $x\in \ZZmum$ prime
to $m$ by the map \beq \chi_{\bullet} (x): \Ifrak_m(\cO_K) \lra
\IC^*,\eeq which is defined on ideals $\pfrak$ prime to $m$ by
sending the prime ideal to the $m$'th root of unity for which \beq
\pfrak~~\mapsto~~\chi_{\pfrak}(x)=x^{\frac{\rmN\pfrak-1}{m}}
(\rmmod~\pfrak).\eeq Using these characters one can define
Jacobi-sums of rank $r$ for any fixed element $a=(a_1,...,a_r)$ by
setting \beq J_a^{(r)}(\pfrak)=(-1)^{r+1} \sum_{\stackrel{u_i\in
\cO_K/\pfrak}{\sum_i u_i=-1 (\rmmod~\pfrak)}}
\chi_{\pfrak}(u_1)^{a_1}\cdots \chi_{\pfrak}(u_r)^{a_r} \eeq for
prime $\pfrak$.  For non-prime ideals $\afrak \subset \cO_K$ the
sum is generalized via prime decomposition $\afrak = \prod_i
\pfrak_i$ and multiplicativity $J_a(\afrak)=\prod_i
J_a(\pfrak_i)$. Hence we can interpret these Jacobi sums as a map
$J^{(r)}$ of rank $r$ \beq J^{(r)}: \Ifrak_m(\ZZmum) \times
(\ZZ/m\ZZ)^r \lra \IC^*, \eeq where $\Ifrak_m$ denotes the ideals
prime to $m$. For fixed $\pfrak$ such Jacobi sums define
characters on the group $(\ZZ/m\ZZ)^r$. It can be shown that for
fixed $a\in (\ZZ/m\ZZ)^r$ the Jacobi sum $J_a^{(r)}$ evaluated at
principal ideals $(x)$ for $x\equiv 1(\rmmod~m^r)$ is of the form
$x^{S(a)}$, where \beq S(a) =
\sum_{\stackrel{(\ell,m)=1}{\ell~\rmmod~m}}\left[\sum_{i=1}^r
\left< \frac{\ell a_i}{m}\right>\right]\si_{-\ell}^{-1},\eeq where
$<x>$ denotes the fractional part of $x$ and $[x]$ describes the
integer part of $x$.

We therefore see that the Hasse-Weil L-function is in fact a
product of functions each of which is determined by a Hecke
character defined by a Jacobi sum that is determined by a prime
ideal in the cyclotomic field $\IQmum$. In the case of the quintic
hypersurface we derive in this way the fusion field from the
arithmetic structure of the defining variety reduced to a finite
field. To summarize, we have seen that the fusion field of the
underlying conformal field theory is precisely that number field
which is determined when the cohomological Hasse-Weil L-function
is interpreted as the Hecke L-function associated to an algebraic
number field.

\section{Quantum dimensions and class number}

At this point we have identified the fusion field with the number
field associated to the Hecke L-function, which in turn is
associated to the Hasse-Weil L-function. In order to extend the
dictionary between conformal field theoretic quantities and number
theoretic quantities we can ask whether we can also obtain a
number theoretic interpretation for the quantum numbers which take
values in the fusion field. This amounts to the question whether
these quantum numbers form some particular substructure of the
cyclotomic field which can be singled out for purely number
theoretic reason.

In order to see that this is indeed possible consider first the
so-called class number $h$ of a cyclotomic field  $\IQmum$. This
number plays an essential role in illuminating the structure of
general algebraic number fields, but our focus will be on the
cyclotomic fields \cite{st79} \cite{r01}. The class number of an
algebraic number field $K$ can be defined as the number of ideal
classes formed by the following equivalence relation. Two ideals
are considered equivalent if they can be made identical by the
multiplication of principal ideals, the latter being generated by
single algebraic numbers. The set of such ideal classes forms a
finite group whose order is defined to be the class number. Put
slightly differently we can think about the ideal class group
$\rmCl(\cO_K)$ as the quotient of the group of fractional ideals
\beq \cF = \left\{\frac{\afrak_1}{\afrak_2}~{\Big |}~ \afrak_1,
\afrak_2 \subset \cO_K, ~\afrak_1, \afrak_2 \neq 0 \right\} \eeq
of the ring $\cO_K$ of algebraic integers in $K$ by the group of
non-zero principal ideals \beq \cP = \left\{(a) = a\cO_K~|~a \in
K^{\times}\right\}. \eeq The importance of the class number $h(K)
= \# \rmCl(\cO_K)$ derives from the fact that it measures how
close a number field comes to unique prime factorization. As an
example consider the rational numbers $\IQ$. Its ring of integers
$\ZZ$ is a principal ring, hence its class number is unity.
Furthermore we have unique prime factorization in $\ZZ$. It turns
out that analogous relations hold in general for algebraic number
fields, i.e. the class number $h(K)$ is unity if and only if there
is unique factorization.

Of essence for the very concept of unique prime factorization are
the so-called units of the field. Such units generalize the units
in the ring of rational integers $\{\pm 1\}\subset \ZZ$, and are
algebraic integers whose inverses are again algebraic integers.
Unique factorization is always meant modulo units. The set of
units in an algebraic field $K$ defines a subgroup $U\subset K$.

Coming back to the cyclotomic fields of interest in the present
context, one finds that the class number of these fields exhibit a
composite structure. It is known that this class number is always
a composite number \beq h(\IQmup) = h^+h^*,\eeq where $h^+$ itself
is the class number of the real cyclotomic subfield of $\IQmup$.

Putting together the concepts introduced in the above paragraphs
provides a number theoretic identification of the quantum
dimensions. Namely, it turns out that the class number $h^+$ of
the real subfield of $\IQmum$ admits an interpretation as the
index of a subgroup within the maximal real subfield in the
cyclotomic field. Let $2<p \neq 2 (\rmmod~4)$ be the conductor of
$\IQ(\mu_p)$ and denote by $U^+$ the group of positive real units
in $\IQ(\mu_p)$, i.e. the positive real invertible elements in the
cyclotomic field. Let further denote $U_c^{+}$ the subgroup
spanned by the cyclotomic units $\theta_r$ within $U^{+}$ \beq
\theta_r = \left|\frac{1-\xi^r}{1-\xi}\right| =
\frac{\sin\frac{r\pi}{p}}{\sin\frac{\pi}{p}}. \eeq Then \beq
h^+=2^{-b} [U^{+}:U_c^{+}], \eeq where $b=0$ if the number $f$ of
prime factors is unity, and $b=2^{f-2}+1-f$ if $f>1$ \cite{s78}.
This provides an identification of the quantum dimensions within
the fusion field.

One can further show that the class number $h^+$ of the real
subfield of $\IQmup$ is in fact partially constructed by the
quantum dimensions. Let $p$ be an odd prime and denote by $\si_j$
the elements $\xi_p \mapsto \xi_p^{g^j}$ of the Galois group
$\rmgal(\IQmup/\IQ)$ of $\IQmup$ generated by a primitive root
modulo $p$.  Then \beq h^+ = \frac{2^{\frac{p-3}{2}}\Delta}{R},
\eeq where the determinant $\Delta$ is constructed from the
quantum dimensions as
 \beq \Delta ~=~\left|\rmdet\left(
\si_j (\theta_k) \right)_{\stackrel{2\leq k\leq (p-1)/2}{0\leq
j\leq (p-3)/2}} \right|.\eeq

The regulator $R$ can be viewed as the volume of the logarithmic
image of a fundamental system of units. It was shown by Dirichlet
that the group $U$ of units in an algebraic number field of degree
$[K:\IQ]=r_1+2r_2$ takes the form \beq U \cong \mu
\prod_{i=1}^{r_1+r_2-1} G_i,\eeq where $\mu$ is the group of roots
of unity, each $G_i$ is a group of infinite order, and $r_1$
($r_2$) denotes the number of real (complex) embeddings of the
field $K$. Hence every unit $u\in U$ can be written in the form
$u=\a \prod_{r=1}^r \e_i$, where $\a \in \mu$ and
$\{\e_i\}_{i=1,...,r=r_1+r_2-1}$ is called a fundamental system of
units. It is useful to translate the multiplicative structure of
the units into an additive framework via the regulator map \beq r:
U \lra \IR^{r_1+r_2} \eeq defined by \beq r(u) =
\left(\ln|\rho_1(u)|,\dots,
\ln|\rho_{r_1}(u)|,\ln|\rho_{r_1+1}(u)|^2,\dots,
\ln|\rho_{r_1+r_2}(u)|^2\right),\eeq where
$\{\rho_i\}_{i=1,\dots,r_1}$ are the real embeddings and
$\{\rho_{r_1+j}\}_{j=1,\dots,r_2}$ are the complex embeddings of
$K$. The regulator $R$ then is defined as \beq R=\rmdet \left( a_i
\ln |\rho_i(\e_j)|\right)_{\stackrel{1\leq i\leq r_1+r_2}{1\leq
j\leq r_1+r_2-1}} \eeq with $a_i=1$ for the real embeddings
$i=1,...,r_1$ and $a_i=2$ for the complex embeddings
$i=r_1+1,...,r_1+r_2$. The regulator is independent of the choice
of the fundamental system of units.

The results described above provide an example where the class
number of an algebraic number field acquires physical
significance. This is not without precedence.
 Recently the class number of the fields of definition
of certain arithmetic black hole attractor varieties have been
interpreted as the number of U-duality classes of black holes with
the same area \cite{m98}.

To summarize, we see that a further entry in our dictionary is
provided by the identification of the quantum dimensions of the
fusion field with the real cyclotomic units of the field that is
determined by recognizing the Hasse-Weil L-function as the Hecke
L-function of an algebraic number field.

 \vskip .3truein

{\bf Acknowledgement} \hfill \break This work has been done over a
number years at several institutions. I'm grateful to Werner Nahm
and his group at the University of Bonn for discussions during its
first period. It is a pleasure to thank Alan Adolphson, Monika
Lynker, Chad Schoen, John Stroyls, John Tate, Alexandre Varchenko,
 Katrin Wendland, and in particular Vipul Periwal, for discussions.
 Parts of this work were
completed  at UT Austin and during the Duality Workshop at the
Institute for Theoretical Physics in Santa Barbara. I'm grateful
to the Theory Group at Austin and the ITP for hospitality. This
research was supported in part by NATO under grant CRG 9710045, an
ITP scholarship, and the National Science Foundation under grant
No. PHY99-07949.


\begin{thebibliography}{9}
\bibitem{g87} D.Gepner, {\it Spacetime Supersymmetry in Compactified String
Theory and Superconformal Models}, Nucl.Phys. {\bf
B296}(1988)757;\\
{\it String Theory and Calabi-Yau Manifolds: The Three Generation
Case}, preprint Dec. 1987, hep-th/9301089
\bibitem{ls90} M.Lynker and R.Schimmrigk, {\it ADE Quantum Calabi-Yau Manifolds},
 Nucl.Phys. {\bf B339} (1990) 121
\bibitem{fkss90} J.Fuchs, A.Klemm, C.Scheich and M.G.Schmidt,
{\it Spectra and Symmetries of Gepner Models Compared to
Calabi-Yau Manifolds}, Ann.Phys. {\bf 204}(1990)1
\bibitem{chsw85} P.Candelas, G.Horowitz, A.Strominger and
E.Witten, {\it Vaccum Configurations for Superstrings}, Nucl.Phys.
{\bf B258}(1985)46
\bibitem{s87} R.Schimmrigk, {\it A New Construction of a
Three Generation Calabi-Yau Manifold}, Phys.Lett. {\bf B193}
(1987) 175
\bibitem{m89} E.Martinec,  {\it Algebraic Geometry and Effective Lagrangians},
Phys.Lett. {\bf B217}(1989)431
\bibitem{vw89} C.Vafa and N.Warner,  {\it Catastrophes and the Classification
of Conformal Field Theories},  Phys.Lett. {\bf B218}(1989)51\
\bibitem{lvw89} W.Lerche, C.Vafa and N.Warner, {\it Chiral Rings in N=2
Superconformal Theories}, Nucl. Phys. {\bf B324} (1989) 427
\bibitem{v88} C. Vafa, {\it String Vacua and Orbifoldized L-G
Vacua}, Mod.Phys.Lett. {\bf A4}(1989)1169
\bibitem{w92} E.Witten, {\it Phases of N=2 Theories in Two in Two Dimensions},
 Nucl. Phys. {\bf B403} (1993) 159, hep-th/9301042
\bibitem{s95} R. Schimmrigk, Lecture at Bonn University, 1995; \\
 Lecture at the {\it Workshop on Arithmetic, Geometry and Physics
 around Calabi-Yau Manifolds and Mirror Symmetry}, J.D. Lewis and
 N. Yui, 2001
\bibitem{m98} G. Moore, {\it Attractors and Arithmetic},
hep-th/9807056; {\it Arithmetic and Attractors}, hep-th/9807087
\bibitem{cdv00} P. Candelas, X. de la Ossa and F. Rodriguez-Villegas, {\it
 Calabi-Yau Manifolds over Finite Fields I}, hep-th/0012233; \\
  P. Candelas, X. de la Ossa and F. Rodriques-Villegas, in
  preparation
\bibitem{ak94} A.Kirillov, {\it Dilogarithm Identities}, Progr. Theor. Phys.
Suppl {\bf 118} (1995) 61, hep-th/9408113
\bibitem{dg91} J. de Boer and J. Goeree, {\it Markov Traces and
II(1) Factors in Conformal Field Theories}, Commun. Math. Phys.
{\bf 139} (1991) 267
\bibitem{w49} A.Weil, {\it Number of solutions of equations in finite fields},
Bull.Am.Math.Soc. {\bf 55}(1949)497; \\
G\"{o}tt.Nachr. {\bf 1}(1974)14
\bibitem{d74} P.Deligne,  {\it La conjecture  de Weil I},
Publ.Math. IHES {\bf 43}(1974)273
\bibitem{k94} S.L.Kleiman,  {\it The standard conjectures},
Proc. Symp. Pure Math. {\bf 55} (1994) 3
\bibitem{bd60} B.Dwork, {\it On the rationality of the zeta
function of an algebraic variety}, Amer.J.Math. {\bf 82}(1960)631
\bibitem{h18} E. Hecke, {\it Eine neue Art von Zetafunktionen und
ihre Beziehungen zur Verteilung der Primzahlen}, Math. Z. {\bf 1}(1918)357;\\
{\it Eine neue Art von Zetafunktionen und ihre Beziehungen zur
Verteilung der Primzahlen. Zweite Mitteilung}, Math. Z. {\bf 6}
(1920)11
\bibitem{d77} P.Deligne, {\it Cohomologie \'etale}, Springer Verlag LNM 569, 1977
\bibitem{st79} I.Stewart and D.Tall, {\it Algebraic Number
Theory}, Chapman and Hall, 1979
\bibitem{r01} P. Ribenboim, {\it Classical Theory of Algebraic Numbers},
Springer Verlag, 2001
\bibitem{s78} W. Sinnott, {\it On the Stickelberger Ideal and the Circular
Units of a Cyclotomic Field}, Ann. Math. {\bf 108} (1978) 107
\end{thebibliography}
\end{document}